\documentclass[
    ,final            
  ]
  {aipproc}

\layoutstyle{6x9}
\begin{document}

\title{A search for wide (sub)stellar companions around extrasolar planet host stars}

\author{M.~Mugrauer}{
  address={AIU, Schillerg\"asschen 2-3, 07745 Jena}
}

\author{R.~Neuh\"auser}
{
 address={AIU, Schillerg\"asschen 2-3, 07745 Jena, Germany}}

\author{T.~Mazeh}{
  address={Tel Aviv University, P.O. Box 39040, 69978 Tel Aviv, Israel}
}

\author{M.~Fern\'andez}{
  address={IAA, Apdo. Correos 3004, 18080 Granada, Spain}
}

\author{E.~Guenther}{
  address={TLS, Sternwarte 5, 07778 Tautenburg, Germany}
}

\begin{abstract}
We present an overview of our ongoing systematic search for wide (sub)stellar companions around the
stars known to host rad-vel planets. By using a relatively large field of view and going very deep,
our survey can find all directly detectable stellar and massive brown dwarf companions
(m$>$40\,$M_{Jup}$) within a 1000\,AU orbit.
\end{abstract}

\maketitle
\section{Introduction}
Circumstellar disks are discovered with sizes up to 1000\,AU and also binary stars with the same
comparable separations are known. Because the formation of stars and brown dwarfs seems to follow a
similar scheme (fragmentation of large gas clouds) substellar objects may indeed reside in that
distance around stars hosting rad-vel planets. Adaptive optics search programs to find very close
companions around those stars already exist, but they leave out an interesting regime of objects,
namely the wide companions because of a too small field of view (e.g. Patience et al. 2002).

As of October 2003, more than one hundred extra-solar planets were discovered. Many of those have
extremely close orbits which could be explained by a migration process in the early history of the
system. During this migration angular momentum is transferred from the inner part of the accretion
disk to its outer border. A wide companion can cut off the disk and be a sink for the lost angular
momentum. Furthermore theories predict that wide companions can induce rapid instability in disks
which otherwise would be stable, hence they could have a strong influence on the planet formation
and on the longtime evolution of planetary orbits.

Actually, some extrasolar planets were found to reside in binary stellar systems. Those few cases
are intriguing, and might exhibit some statistically different features than the planets around
single stars (Zucker \& Mazeh 2002). This could be a first hint about an interaction between the
(sub)stellar wide companions and the extrasolar planets. Nevertheless, the whole sample of
extrasolar planetary systems has not been surveyed completely for wide companions with sensitive IR
cameras that are able to find faint low-mass companions. For this reason we have started in 2001 a
systematic deep imaging of all the stars known to harbor planets, in order to look for faint
companions in wide orbits. The companionship of those faint objects can be established only by
follow up observations which will detect common proper motion with the nearby star that host the
planet.

To find all the companion-candidates around the planet hosting stars we secure deep IR images,
obtained with the IR cameras SOFI at the 3.58\,m NTT and UFTI at the 3.8\,m UKIRT, with detection
limit of H\,$\sim$\,19.5\,mag. To detect common proper motion we obtain two images about one year
apart. We also make use of the 2 micron all sky survey (2MASS) images, which were taken several
years before our exposures. However, the limit of 2MASS is H\,$\sim$\,15\,mag (Cutri et al. 2003),
and the proper motion of fainter objects needs to be measured only by our two images.

\section{Astrometry - an effective way to find companions}

Astrometry is a very effective tool to find unknown wide companions. All stars hosting planetary
systems are bright and therefore are listed in the Hipparcos catalogue, hence their proper motions,
which are relatively high ($\sim$~200\,mas per year), are known with an accuracy of a few
milliarcsecs (mas). With the IR cameras --- UFTI/UKIRT and SOFI/NTT, which have pixel scales of
91\,mas and 144\,mas, respectively, one year difference means a shift of the photocenter by 1.5 to
2 pixel. This is easily detectable. Only two observations are necessary to distinguish real
companions from background stars. A real companion is bound to its host star and therefore they
form common proper motion pair. Co-moving companions stand out with non varying separation, whereas
the separation between background stars and the target star changes according to the well known
proper motion of the target star. The orbital motion of the companion around its host star can be
neglected, because the motion of wide companions with orbital radii larger than 100\,AU is
generally much smaller than the proper motion.

\begin{figure}
\includegraphics[height=.3\textheight]{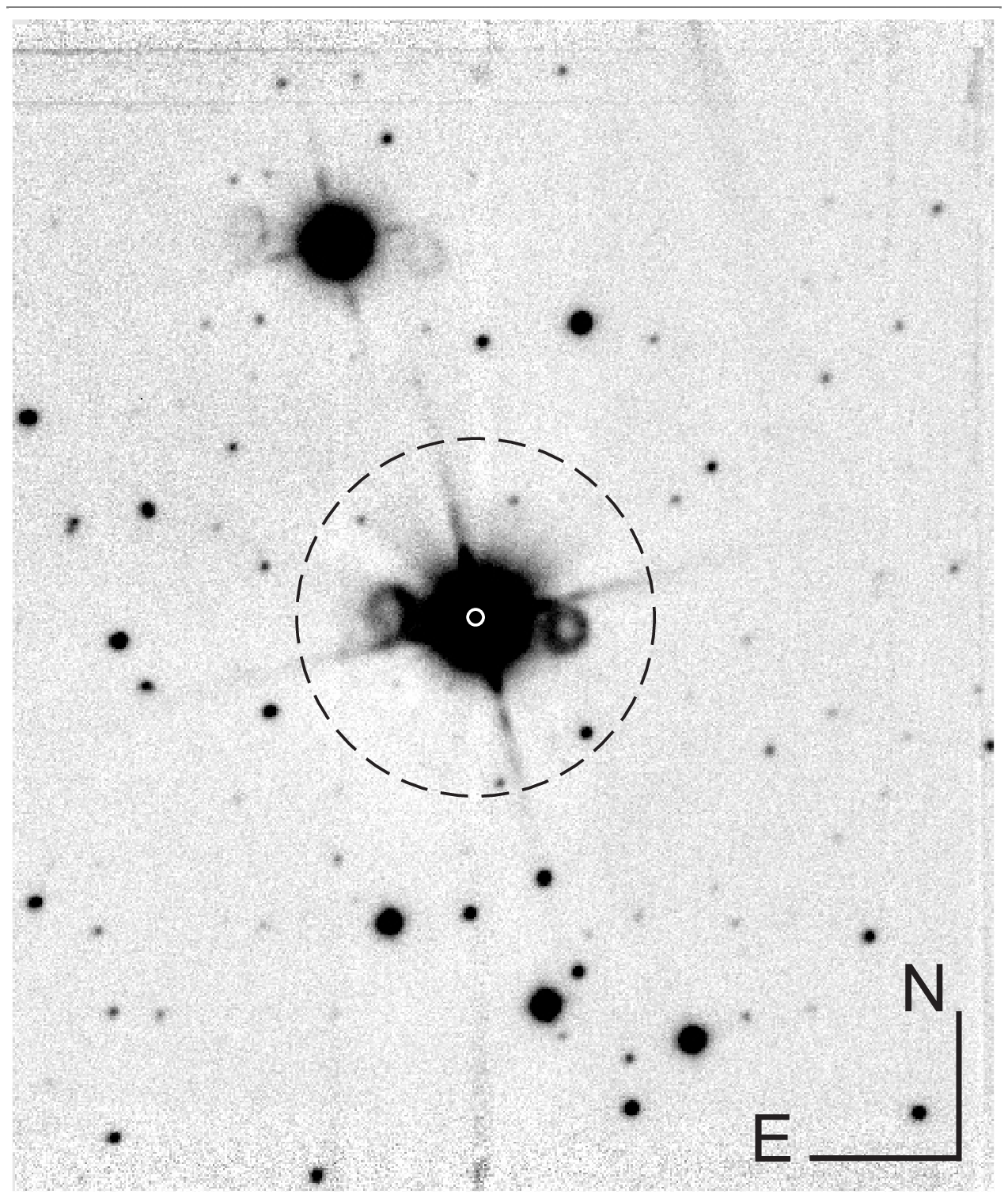}\hspace{1cm}\includegraphics[height=.3\textheight]{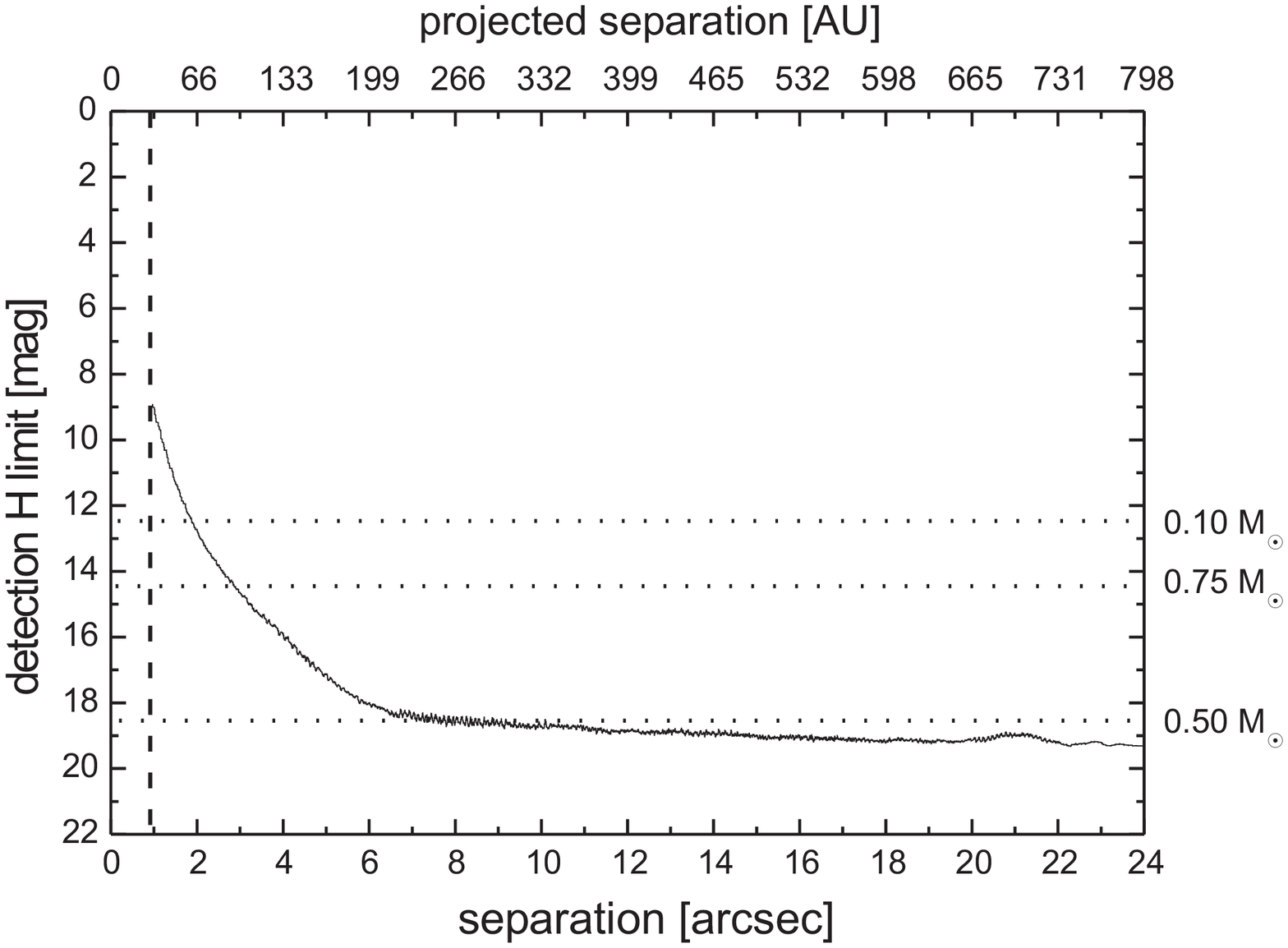}
  \caption{NTT H band image of the field of the rad-vel planet host star HD\,37124 which was taken in Dec. 2002. The
  averaged FWHM is $\sim$0.8\,arcsec. Several companion-candidates are detected close to the rad-vel
  planet host star with magnitudes down to H\,=\,19.5\,mag. Inside 1\,arcsec saturation occurs (see the white
  small circle in the left image and the dashed line in the right panel). The right panel shows the detection
  limit for a range of separations. The field of view of the derived detection limits is marked in the left
  image as dashed black circle. The upper x-axis shows the projected separation in AU. The right y-axis scales
  in companion mass. We used Baraffe et al. (2003) COND models for the magnitude-mass transformation, assuming an
  age of 5\,Gyrs for the system.}
\end{figure}

An example of the results of our astrometric search campaign is illustrated in fig.1 \& 2. We
observed the star HD\,37124 (for planet data see Vogt et al. 2002) with SOFI/NTT in Dec. 2001 and
Dec. 2002. Total integration time was 10\,min in H in both exposures. The SOFI/NTT field of view is
144\,arcsecs, hence companions with maximum projected separation of 2200\,AU can be detected. Close
to the bright star the detector becomes non-linear and saturation occurs (see the white circle in
fig.\,1 left and the dashed line in fig.\,1 right). In this region the detection of any companion
is impossible. To avoid saturation at large parts of the detectors we choose as short integration
time as possible. Many of those short integrated images have to be superimposed to reach a high
signal to noise.

In the next step we have derived the detection limit of our observations by measuring the local
noise in the images for a range of separations to the rad-vel planet host star. The detection
sensitivity is increasing for larger separations due to a lower local photon noise. In the right
panel of fig.\,1 we have plotted the detection limit for a range of angular distances to HD\,37124.
Only a small fraction of the whole SOFI/NTT field of view (147\,arcsecs) is shown here. Beyond
24\,arcsecs the image is background limited and the detection limit is constant, around 19.5\,mag
(see the dashed black circle in fig.\,1, left panel). In this region brown dwarf companions with
$m$\,$>$\,40$M_{Jup}$ are detectable. All stellar companions (m\,$>$78\,M$_{Jup}$) can be detected
outside $\sim$3\,arcsecs, with projected separation of $\sim$100\,AU. We use evolutionary models to
derive the apparent magnitudes of objects with different masses at the distance of HD\,37124 for an
age of 5\,Gyrs.

The proper motion of all detected objects around the star can be measured by using relative
astrometry. Only the star itself has a non-negligible proper motion which is consistent with the
Hipparcos proper motion data (see the white box in fig.2). All other detected objects are clearly
not co-moving, hence they are all non-moving background stars. Due to the results of our relative
astrometry and the derived detection limit we can conclude that there is no further stellar
companion around HD\,37124 between 100 and 2200\,AU (3...73\,arcsecs).

So far nearly all rad-vel planet host stars were observed in first epoch and second epoch follow-up
observations are on the way. Several new wide companions were detected which we will be published
soon.

\begin{figure}
  \includegraphics[height=.3\textheight]{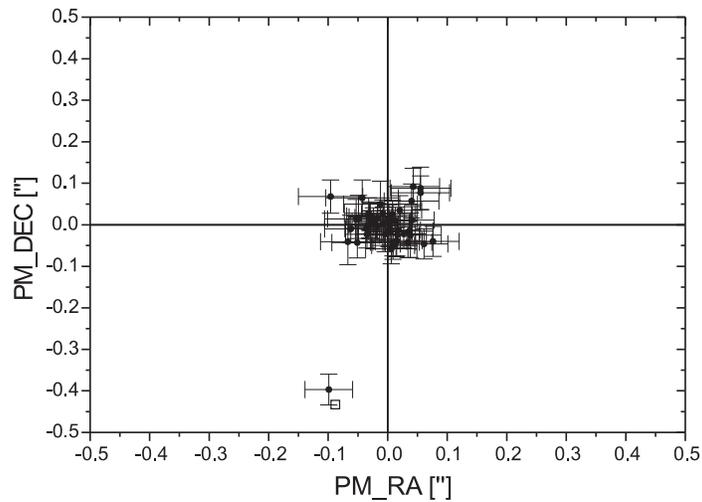} \caption{Measured proper motion
  of all objects in fig.\,1, obtained by comparing two SOFI/NTT H band images (relative astrometry)
  taken with an epoch difference of one year. The proper motion of HD\,37124 is well known
  from Hipparcos measurements, and is presented here by the small squared box. All
  detected objects have proper motions negligible within the astrometry precision, hence
  they are none-moving background stars. Only HD\,37124 itself is moving with the
  expected proper motion.}
\end{figure}

\begin{theacknowledgments}
We would like to thank the technical staff of NTT and UKIRT for all their help and assistance in
carrying out the observations. The United Kingdom Infrared Telescope (UKIRT) is operated by the
Joint Astronomy Centre on behalf of the U.K. Particle Physics and Astronomy Research Council. This
publication made use of data products from the Two Micron All Sky Survey, which is a joint project
of the University of Massachusetts and the Infrared Processing and Analysis Center/California
Institute of Technology, funded by the National Aeronautics and Space Administration and the
National Science Foundation. We have used the SIMBAD database, operated at CDS, Strasbourg, France.
Finally we would like to thank Christopher Broeg, Andreas Seifahrt and Alexander Szameit for
carrying out some of the observations.
\end{theacknowledgments}

\bibliographystyle{aipproc}

\end{document}